\documentclass[11pt,a4paper]{article}

\usepackage[utf8]{inputenc}
\usepackage[T1]{fontenc}
\usepackage{lmodern}
\usepackage[margin=25mm]{geometry}
\usepackage{microtype}
\usepackage{amsmath}
\usepackage{graphicx}
\usepackage{booktabs}
\usepackage[font=small,labelfont=bf]{caption}
\usepackage[sort&compress,numbers]{natbib}
\usepackage{fancyhdr}
\usepackage{xcolor}
\usepackage[hyphens]{url}
\usepackage[colorlinks=true,allcolors=blue!60!black]{hyperref}
\hypersetup{
  pdftitle={What is non-hyperbolicity, that we should be mindful of it? Noise as a constituent of chaos},
  pdfauthor={Igor A. Khovanov},
  pdfkeywords={Chaos; noise; non-hyperbolicity; numerical noise; predictability; large deviations; data-driven models}
}
\newcommand{\rr}{\raggedright\hspace{0pt}}

\title{\bfseries\Large What is non-hyperbolicity, that we should be mindful of it?\\[4pt]
  \large Noise as a constituent of chaos}
\author{Igor A. Khovanov\thanks{\href{mailto:i.khovanov@warwick.ac.uk}{i.khovanov@warwick.ac.uk};
  ORCID: \href{https://orcid.org/0000-0002-0372-8457}{0000-0002-0372-8457}.
  Preprint of an article submitted for consideration in \emph{Fluctuation and Noise Letters}
  (\url{https://www.worldscientific.com/worldscinet/fnl}).}\\[6pt]
  \normalsize\itshape School of Engineering, University of Warwick,\\
  \normalsize\itshape Coventry CV4 7AL, United Kingdom}
\date{}

\begin{document}

\maketitle
\thispagestyle{plain}

\begin{abstract}
\noindent
No chaotic trajectory has ever been observed free from perturbation: in experiments, it arises from thermal and environmental fluctuations, while in numerical simulations, it stems from truncation and round-off errors. I argue that this ubiquitous presence of noise carries three fundamental consequences for the study of chaotic dynamical systems.
First, a typical chaotic trajectory, whether computed or measured, has the status of a realization of a stochastic process rather than of an exact deterministic solution, and must be treated as such. This profoundly affects data-driven machine-learning models. Second, the noisy system itself becomes the proper object of study, where the noise intensity sets a fundamental resolution limit below which the fine structure of an attractor ceases to have physical meaning. Third, the dynamical role of noise depends entirely on the question being asked. While noise obscures non-hyperbolicity in bulk statistics at any fixed noise level, it actively exposes specific underlying structures of the attractor through rare large deviations and the precise manner in which predictability is lost.
I present evidence supporting this framework for quasi-hyperbolic and non-hyperbolic attractors. Finally, I examine the implications for numerical practice and data-driven modeling, and outline open problems.

\end{abstract}

\section{The Question}
\label{sec:question}

In 1989, Joseph Ford inquired, ``What is \emph{chaos}, that we should be mindful of it?''~\cite{Ford89} Decades later, the significance of chaos theory is self-evident and no longer requires justification. The question I want to ask is narrower, and it concerns a word that many who work with chaotic systems seldom meet: what is \emph{non-hyperbolicity}, that we should be mindful of it? The theory has been developed mainly for hyperbolic chaos. The attractors of real systems and models, with few exceptions, are quasi-attractors (non-hyperbolic attractors).  Additionally, beyond the theory, real chaos always includes random perturbations, such as unavoidable device noise in experiments and truncation and round-off errors in numerical models.  Dorfman stated the alternative: random perturbations may be ``a serious complicating factor,'' or they may make ``non-hyperbolic systems behave as if they were hyperbolic''~\cite{Dorfman99}.  More than twenty-five years later, there is still no consensus on how to link chaos theory to reality.

This question is linked to a more fundamental one, older than the study of chaos. In the 1930s, Andronov's school, responding to the demand for mathematical tools for electrical circuits, formulated two types of robustness for any model of a real device: robustness against fluctuations, to be treated statistically~\cite{PAV33}, and robustness against small changes in the model itself (the model should be \emph{rough})~\cite{AP37}. Only model properties that satisfied both were to count as physical. These two robustness criteria were considered natural and were implicitly assumed, since typical models for regular motion meet both. The only explicit formulation of these two criteria, presented as self-evident, appears in a footnote to the introduction of the second edition of the book by Andronov, Vitt and Khaikin~\cite{AVH_book}. 

For chaotic motion, in contrast to regular motion, where typical models meet both demands, the criteria are generally not met. Moreover, they have been pursued separately: the first in stochastic dynamics and the second in the theory of structural stability.  The first avenue is a success story: the original paper was recognized as its source in the famous three-volume collection \emph{Noise in Nonlinear Dynamical Systems}~\cite{NNDS3vols}, edited by Frank Moss and Peter V. E. McClintock, which established stochastic nonlinear dynamics as a fruitful interdisciplinary research direction. The second has been more complicated and has developed alongside advances in mathematical rigor. It led to the discovery of Newhouse regions~\cite{Newhouse79} and Smale's statement~\cite{Smale} that a typical chaotic system is structurally unstable, not rough, and culminated in Gonchenko, Shil'nikov and Turaev's conclusion~\cite{GST93}: near a homoclinic tangency, arbitrarily degenerate phenomena occur, so a complete qualitative description of a generic chaotic system is impossible. The latter remains a niche result that is rarely appreciated. 

Here, I argue that for chaotic systems both criteria must be taken together. This journal has played a part in the matter from the start: its first volume includes a letter showing that noise-induced escape from the Lorenz attractor is governed by the attractor's structure where hyperbolicity fails~\cite{FNL01}. In this opinion paper, I outline how the matter extends beyond that result. Section~\ref{sec:nonoisefree} argues that there is no noise-free chaos, in the laboratory or in the computer. Section~\ref{sec:realisation} draws the consequences for the status of a trajectory. Section~\ref{sec:types} recalls the types of chaotic attractors and explains why the typical one is not rough. Section~\ref{sec:faces} gives my answer to Dorfman's alternative: the effect of noise depends on the question asked, and differs between bulk statistics, rare large deviations and predictability. Section~\ref{sec:consequences} draws consequences for numerical practice, for the treatment of model error and for models learned from data. Section~\ref{sec:open} lists open problems.

\section{There Is No Noise-Free Chaos}
\label{sec:nonoisefree}

\begin{figure}[!t]
\centering
\includegraphics[width=0.9\textwidth]{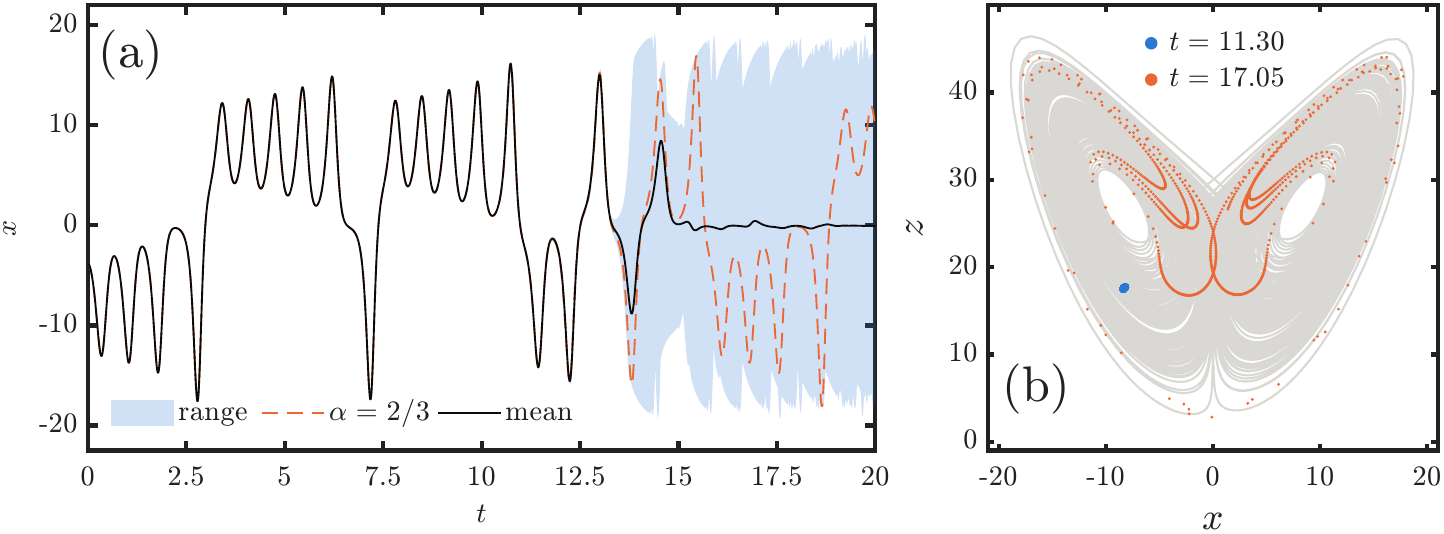}\\
\includegraphics[width=0.9\textwidth]{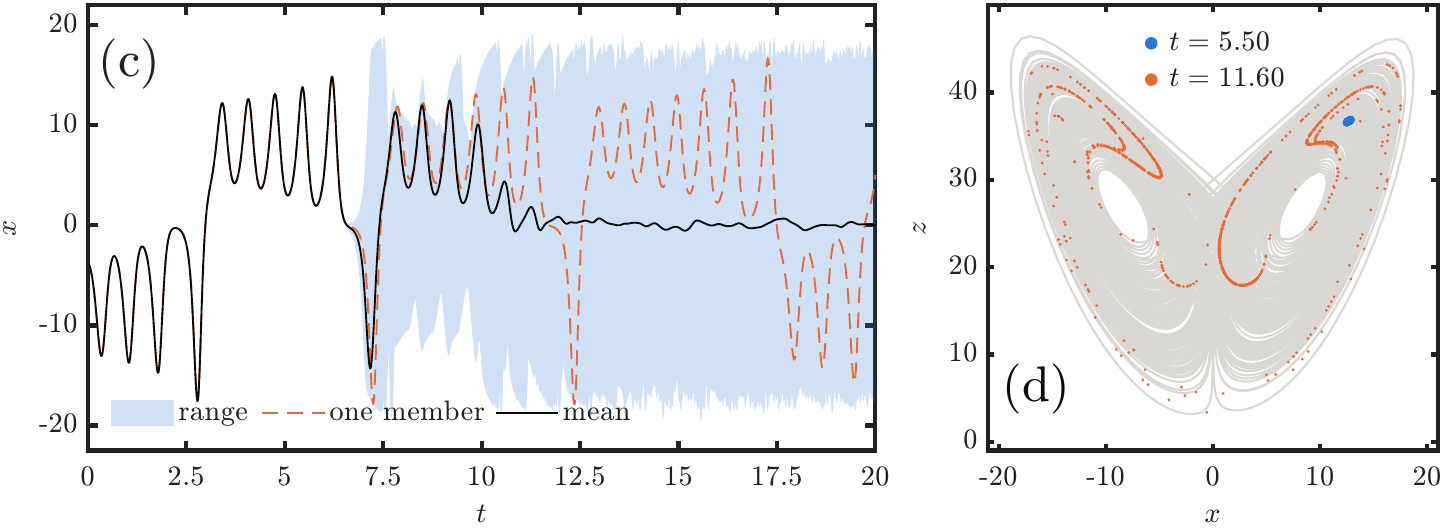}
\caption{Lorenz system at the standard parameter values: $\sigma=10$, $r=28$ and $b=8/3$, which correspond to the quasi-hyperbolic attractor. An ensemble of numerical schemes (panels (a) and (b), top) and a noise ensemble (panels (c) and (d), bottom), computed from the same initial state, with the same step size $h=0.001$, and 1000 members each. 
Top, (a)--(b): explicit second-order Runge--Kutta schemes, $k_1=f(x_n)$, $k_2=f(x_n+\alpha h k_1)$, $x_{n+1}=x_n+h[(1-\frac{1}{2\alpha})k_1+\frac{1}{2\alpha}k_2]$, with $\alpha$ equally spaced in $[\frac12,1]$ (midpoint scheme, $\alpha=1/2$, to Heun's scheme, $\alpha=1$) were used. The ensemble member shown with the red dashed line corresponds to $\alpha=\frac23$ (Ralston's scheme). 
Bottom, (c)--(d): the stochastic fourth-order scheme of Ref.~\cite{Khovanov21} with noise intensity $D=10^{-6}$ was used. An ensemble member is shown with the red dashed line. In panels (a) and (c), the range of the coordinate $x$ over the ensemble members and their mean are shown by the shaded area and the solid black line, respectively. In panels (b) and (d), all ensemble members are shown in the $(x,z)$ plane at the times indicated in the plots.  }
\label{fig:lorenz}
\end{figure}

In experiments, noise is always present. A dissipative system is coupled to an environment, and the coupling that produces dissipation also produces fluctuations~\cite{Weiss99}. An electronic circuit, a laser or a nanomechanical resonator in a chaotic regime is a noisy system, and nobody has measured a chaotic trajectory without a noise contribution. It is less often acknowledged that the same is true in computer simulations, which is where most chaos research is done. A numerical solution of an ordinary (or partial) differential equation $\dot{x}=f(x)$ is not a trajectory of that equation. It is a pseudo-orbit: a sequence of points in which every step departs from the true flow by a small error. The error has two sources: the numerical method and round-off.  The first is the truncation error of the scheme's Taylor expansion, $x(t+h)=x(t)+\sum_{i=1}^{p}\beta_i h^{i}f^{(i)}(x)+O(h^{p+1})$, where $h$ is the step size, $\beta_i$ are the scheme's coefficients, $p$ is the order of the scheme, $f^{(i)}$ is the $i$-th derivative along the flow and the last term is the truncation error~\cite{HLW06}. Truncation is therefore a perturbation of the \emph{true} model, and a different one for each scheme.  The second, round-off error, perturbs the state. It is deterministic and reproducible, but state-dependent and, in general, computer-dependent, and, for practical purposes, pseudo-random. Neither is noise in the sense of a stochastic process; together they constitute ``numerical noise.''  Numerical noise can be reduced but not removed in chaotic dynamics because of the divergence of nearby chaotic trajectories (the positive Lyapunov exponent). In Ref.~\cite{Khovanov21}, it was shown that, in the Lorenz system, the solutions computed with two fourth-order schemes and the same step size are initially close but quickly diverge. Figure~\ref{fig:lorenz}(a) shows a family of second-order schemes~\cite{HLW06} doing the same: 1000 schemes start from the same initial condition, agree to the eye until $t\approx 12.7$, and then spread over the attractor. Statistical measures such as Lyapunov exponents, invariant densities and power spectra are computed from runs far longer than the divergence time $t$. All of them are therefore statistics of a perturbed system.

The standard reassurance for a computed trajectory is the shadowing property, which guarantees that close to the computed pseudo-orbit there is a true trajectory with a slightly different initial condition~\cite{HYG87,GHYS90}. This property holds for hyperbolic attractors~\cite{Katok95}, which rarely model real systems.  Typical systems have non-hyperbolic attractors, where shadowing holds only for finite time and breaks down near homo- and heteroclinic tangencies~\cite{HYG87,GHYS90}. Moreover, shadowing fails much sooner when the number of unstable directions varies along the attractor~\cite{Dawson94,SGY97}. Two points follow from the above discussion. Whether a computed chaotic trajectory can be trusted is itself a question about hyperbolicity. Even where shadowing holds, it certifies the computed trajectory as \emph{a trajectory} of the system, not as \emph{the solution} of the initial value problem that was posed.

Several comments clarify how this conclusion relates to rigorous results and to chaos research in general.  Computer-assisted proofs~\cite{Tucker02} have established properties of Lorenz (quasi-hyperbolic) attractors rigorously, but they do so by following rigorous enclosures of sets along a trajectory. Multiple-precision integration can deliver a trajectory of a quasi-hyperbolic attractor over a prescribed finite interval~\cite{Liao09}, at a cost in digits that grows linearly with the interval length. However, the interval is finite, and the conclusion stands. In chaos research, the conclusion is implicit: a trajectory is often an illustrative tool, while rigorous results are obtained by other methods, such as the localization of invariant sets and the analysis of their bifurcations~\cite{Chua_nummethods}.

Thus, numerical noise, and therefore a perturbation of the model, cannot be removed. It should be acknowledged explicitly if simulations are to be linked to real systems. In Ref.~\cite{Khovanov21}, I proposed adding an explicit stochastic term to the equations, with an intensity $D$ above the numerical noise level. The uncontrolled, scheme-dependent perturbation is then replaced with one with known statistics that can be chosen to represent the real system's fluctuations, and the step size is tied to a stated noise level instead of being arbitrary. Figure~\ref{fig:lorenz}(c, d), bottom row, shows the prescription at work. I computed the ensemble of trajectories with the same step, from the same state, and plotted them in the same way as for the numerical noise (Fig.~\ref{fig:lorenz}(a, b)); only the perturbation differs. The noise ensemble spreads about six time units earlier than the ensemble of numerical schemes, since the noise intensity is larger than the equivalent intensity (around $10^{-9}$) of the numerical noise in Fig.~\ref{fig:lorenz}(a, b). Figure~\ref{fig:lorenz} thus shows the same picture for both types of noise. In the Lorenz system, the two numerical solutions do not drift apart gradually as one would expect for a hyperbolic attractor. They separate abruptly as the trajectory passes the region where the stable manifold of the saddle equilibrium divides the flow~\cite{Khovanov21}. This same region controls noise-induced escape from the attractor~\cite{FNL01,JETP02}, and it is the non-hyperbolic structure of the Lorenz attractor that studies of the Lorenz system typically ignore. Thus, numerical noise, like physical noise, is amplified at the attractor's non-hyperbolic element.

\section{A Trajectory Is a Realization, Not a Solution}
\label{sec:realisation}

A typical chaotic trajectory should be treated as one treats a realization of a stochastic process. Nobody presents one trajectory of a Langevin equation as its solution, or asks where that trajectory will be at a distant time. For a Langevin equation, one reports its statistics. The same practice is required for a chaotic system. This is a strictly practical claim that focuses exclusively on computable and measurable outcomes, without denying the theoretical existence of deterministic trajectories.

While several elements of this view are well established, the present work extends them in two key directions. Classical ergodic theory already characterizes a chaotic attractor by its invariant measure~\cite{EckmannRuelle85}, and numerical weather prediction transitioned from single-trajectory forecasts to ensembles once individual runs were recognized as mere realizations of a broader distribution~\cite{Epstein69,Leith74}. Similarly, Lai, Grebogi, and Kurths demonstrated that where shadowing fails, deterministic models can yield only statistical information~\cite{LGK99}. Here, I contribute two distinct advances. The first concerns scope: the conclusion of Sec.~\ref{sec:nonoisefree} applies universally to every chaotic system, in numerical simulations as much as in physical experiments. The second concerns the nature of the ensemble itself: whereas a traditional forecasting ensemble samples across initial conditions to represent observational uncertainty, the ensemble of Ref.~\cite{Khovanov21} samples across noise realizations starting from a single, exactly known initial state. This framework quantifies how rapidly the system itself loses memory of a known state under a given noise level. The timescale over which this ensemble diverges defines a state-dependent prediction horizon that cannot be extended by any improvement in initial-state measurement. The horizon also depends on the noise intensity.

Note that the ensemble of schemes in Fig.~\ref{fig:lorenz} makes the same point without any noise in the computation. The choice of $\alpha$ is deterministic and continuous. Every $\alpha$ value defines a correct second-order scheme, and none is better than the others. Beyond the prediction horizon, the trajectory ensemble fills the attractor, as in the noise case. The choice of scheme has therefore become, for all practical purposes, a random variable, and the scheme parameters define the noise level. 

The claim of this section is subject to three caveats. First, a single long trajectory remains a valid sample for stationary statistics, much like a single long realization of an ergodic random process. However, this holds only when the system converges rapidly to a unique stationary state, as is the case for hyperbolic attractors and the Lorenz attractor (Sec.~\ref{sec:types}). Second, the claim breaks down for quasi-attractors (non-hyperbolic attractors) in which multiple attracting sets coexist. At the parameter values examined in Ref.~\cite{Khovanov21}, for example, the Anishchenko--Astakhov oscillator exhibits two chaotic attractors alongside a stable limit cycle. Weak noise induces transitions among these coexisting sets only rarely, so that, over accessible timescales, a single trajectory samples only one set. In such regimes, an ensemble is required to capture stationary statistics as well as predictions. Third, the claim applies only to \emph{typical} trajectories. Other important features of chaotic systems, such as saddle cycles, invariant manifolds, and optimal fluctuational paths, are specific trajectories, or sets of them, with an individual meaning, and Sec.~\ref{sec:faces} returns to them.

This issue becomes particularly acute in contemporary data-driven modeling, where chaotic systems serve as standard benchmarks for the \emph{world models}~\cite{WM} underlying model-based reinforcement learning in robotics and physical agents. For over two decades, standard practice has been to train a machine-learning model of a chaotic system on a single numerical trajectory and evaluate its performance by how long its forecast tracks a single reference trajectory~\cite{JaegerHaas04,Pathak18}. However, a prediction horizon measured in this manner merely records agreement with one specific realization of the numerical noise inherent to a particular integration scheme. In Ref.~\cite{Khovanov21}, when a trained reservoir computer was evaluated against an ensemble of 1000 noisy realizations initialized from the exact same state, the resulting prediction horizon formed a broad distribution. Crucially, the horizon measured against the noise-free reference was atypical, overestimating the average predictive skill. Consequently, prediction horizons should be reported as distributions and benchmarked against the intrinsic spread of the stochastic ensemble, which establishes a fundamental upper bound. While this evidence currently rests on two dynamical systems~\cite{Khovanov21} and a single machine-learning architecture, broader empirical validation is needed.

\section{Typical Chaos Is Not Rough}
\label{sec:types}

Four types of chaotic attractors are known in dissipative systems, differing in how much of Andronov's roughness (structural stability) they retain.

In a \emph{hyperbolic} attractor, the tangent space at every point splits into fixed numbers of uniformly contracting and uniformly expanding directions~\cite{Katok95}. Such an attractor is rough in the full sense: it is structurally stable under parametric perturbation, noisy trajectories are shadowed by true ones for all time, and its statistics vary little under weak noise~\cite{Kifer88}. Yet it is also rare; indeed, an electronic circuit exhibiting a hyperbolic attractor had to be specifically engineered for the purpose~\cite{Kuznetsov05}.

The \emph{Lorenz attractor} exemplifies the second, \emph{quasi-hyperbolic} type. It is hyperbolic except that it contains a saddle equilibrium together with
its manifolds~\cite{Tucker02,Sparrow82}. It is robust, in that it persists under small perturbations and contains no stable periodic orbits, but it is
not structurally stable in the strict sense~\cite{GW79}.

The third type, the \emph{quasi-attractor} (\emph{non-hyperbolic attractor})~\cite{AS83}, is the generic case. It is structurally unstable (not rough), comprising non-trivial hyperbolic sets alongside stable periodic orbits of long period and narrow basins, as well as homoclinic tangencies between the invariant manifolds of its saddle orbits. An arbitrarily small parameter variation destroys existing tangencies while creating new ones, a phenomenon that persists across entire parameter intervals~\cite{Newhouse79}, rendering a complete description of the dynamics impossible in principle~\cite{GST93}.

The fourth type, the \emph{pseudohyperbolic} attractor, which encompasses \emph{wild} attractors~\cite{TS98}, likewise admits homoclinic tangencies, yet no stable periodic orbits can emerge under small perturbations. Its chaos is robust, even though its detailed phase-space structure is not. Concrete examples in continuous-time flows have been established only recently~\cite{GKT21}.

For the first type, noise acts merely as a small perturbation: the deterministic attractor dominates, and weak noise simply blurs its fine structure. For the second type, noise can occasionally probe a rarely visited non-hyperbolic structure~\cite{FNL01}, whereas away from it the hyperbolic picture still applies. For a quasi-attractor, in my view, this framework breaks down. What a numerical simulation or physical experiment reveals as the ``chaotic attractor'' is generated jointly by the deterministic dynamics and the stochastic perturbation. The observed chaos is, in part, created by the noise itself. Noise thus becomes an intrinsic constituent of the object under study rather than an external disturbance to it. The zero-noise limit, in which the intricate deterministic fine structure re-emerges, describes nothing that can ever be observed. Consequently, the question ``what is the attractor?'' must be replaced by ``what is the attractor at this level of noise?''

\section{Two Faces of Noise: It Depends on the Question}
\label{sec:faces}

Dorfman framed the dichotomy sharply: random perturbations may act as a ``serious complicating factor,'' or they may make ``non-hyperbolic systems behave as if they were hyperbolic''~\cite{Dorfman99}. Both outcomes are well documented, leading to the conventional view that the role of noise depends primarily on the system under study. In my view, however, it depends first and foremost on the question being asked. Three distinct questions can be put to a noisy chaotic system: what its typical stationary statistics are, how it executes rare large excursions, and how far ahead its state can be predicted. Noise answers each of them differently.

\begin{table}[!t]
\centering
\caption{What weak noise does to a chaotic attractor, by type of attractor and by
question asked. $D$ is the noise intensity.}
\label{tab:faces}
\footnotesize
\setlength{\tabcolsep}{3pt}
\begin{tabular}{@{}p{0.15\textwidth}p{0.20\textwidth}p{0.20\textwidth}p{0.20\textwidth}p{0.16\textwidth}@{}}
\toprule
\rr Type of attractor & \rr Bulk statistics & \rr Rare large deviations &
\rr Predictability & \rr State of evidence \tabularnewline
\midrule
\rr Hyperbolic~\cite{Kuznetsov05} &
\rr Change little and continuously with $D$~\cite{Kifer88} &
\rr Quasipotential defined for hyperbolic sets~\cite{GHT91}; escape not
studied in a flow &
\rr Shadowing for all time~\cite{Katok95}; horizon set by
the Lyapunov exponent &
\rr Theory mature, mainly for maps; noise in hyperbolic flows barely studied
\tabularnewline\addlinespace[4pt]
\rr Quasi-hyperbolic (Lorenz) &
\rr Change little with $D$~\cite{Anishchenko00} &
\rr Optimal path unique, set by the manifolds of the saddle
equilibrium~\cite{FNL01,JETP02} &
\rr Lyapunov rate, then abrupt loss at the manifold-splitting
region~\cite{Khovanov21} &
\rr One system
\tabularnewline\addlinespace[4pt]
\rr Quasi-attractor &
\rr Regularized; look hyperbolic at given $D$~\cite{Anishchenko00}, but depend
on $D$~\cite{Anishchenko01}; large displacement near
tangencies~\cite{JaegerKantz97,SOY98} &
\rr Path starts at a homoclinic tangency~\cite{KG04} or on a saddle
cycle~\cite{Khovanov00,Khovanov08}; unresolved &
\rr Shadowing finite or absent~\cite{HYG87,GHYS90,Dawson94,SGY97}; error can
stay bounded well beyond the Lyapunov time~\cite{Khovanov21} &
\rr Several maps and oscillators; partial and conflicting
\tabularnewline\addlinespace[4pt]
\rr Wild (pseudo\-hyperbolic)~\cite{TS98,GKT21} &
\rr Chaos robust without noise~\cite{TS98,GKT21}; with noise unknown &
\rr Unknown & \rr Unknown & \rr None
\tabularnewline
\bottomrule
\end{tabular}
\end{table}

The first question concerns the bulk (high-probability core) of the stationary distribution, where weak noise has a simplifying effect. For a hyperbolic attractor, the invariant measure, Lyapunov exponents, and correlation functions vary little and continuously with the noise intensity $D$~\cite{Kifer88}. The Lorenz attractor behaves similarly~\cite{Anishchenko00}. For a quasi-attractor, by contrast, the statistics change substantially: periodic windows disappear, the invariant measure is smoothed, and at any fixed $D$ the system becomes virtually indistinguishable from a hyperbolic one~\cite{Anishchenko00}. The earliest quantitative example is the period-doubling cascade, which noise truncates at a level governed by a universal scaling law~\cite{CNR81,CFH82}. This observation points to a broader principle. The noise intensity sets an effective resolution, and any phase-space structure finer than that scale, such as a narrow periodic window, a sink of long period, or the successive tangencies of a Newhouse region, ceases to be a property of the physical system. At finite $D$, the statistics vary continuously with system parameters even where their zero-noise limits fail to do so. Here, the two demands of the Andronov school merge: roughness must be judged at the scale of the noise. Nevertheless, non-hyperbolicity is merely hidden in the bulk rather than erased. The stationary statistics retain a marked dependence on $D$, the mixing rate shifts~\cite{Anishchenko01}, and near homoclinic tangencies the noisy attractor is displaced by far more than the noise amplitude itself~\cite{JaegerKantz97,SOY98}. A noisy quasi-attractor appears hyperbolic at any single noise level, yet betrays its nature in how it evolves from one level to the next.

The second question concerns the tails (rare, dramatic events). A large deviation from the attractor proceeds along an optimal fluctuational path~\cite{FW84}, and this path does not sample the attractor uniformly. For the Lorenz attractor, the optimal path is unique and is determined by the invariant manifolds of the saddle equilibrium~\cite{FNL01,JETP02}, the sole element where hyperbolicity breaks down. For non-hyperbolic attractors in maps, Kraut and Grebogi find that the path originates at the primary homoclinic tangency closest to the basin boundary~\cite{KG04}. By contrast, for a periodically driven oscillator, the experimentally measured path emanates from a low-period saddle cycle embedded within the attractor~\cite{Khovanov00,Khovanov08}. These two findings have yet to be reconciled, and the distinction is fundamental: a homoclinic tangency is a locus where hyperbolicity fails, whereas a saddle cycle is itself hyperbolic. Yet both mechanisms agree on one essential point: rare events select a sparse subset of phase-space structures that the bulk statistics simply average away.

The third question concerns how far ahead the state can be predicted. The conventional answer is the Lyapunov time, set by the largest Lyapunov exponent. In the two systems examined in Ref.~\cite{Khovanov21}, this timescale gives the right order of magnitude for the moment when an ensemble of noisy trajectories, initialized from a single point, first spreads appreciably. However, it fails to describe the manner in which predictability is lost. In the Lorenz system, the ensemble remains compact before spreading across the attractor abruptly as it passes through the region where the invariant manifolds of the saddle equilibrium divide the flow. In the Anishchenko--Astakhov oscillator, which possesses a quasi-attractor, the spread initially grows at the Lyapunov rate but subsequently oscillates without reaching the full scale of the attractor for roughly ten times the Lyapunov estimate. Admittedly, two systems do not establish a general rule. Nevertheless, they demonstrate that a global exponent, being an average over the attractor, cannot reveal where predictability breaks down or how completely it is lost. Shadowing theory makes the complementary point from the opposite perspective: the duration for which a noisy trajectory remains close to a true deterministic one is governed by homoclinic tangencies and fluctuations in the number of unstable directions, rather than by the largest Lyapunov exponent~\cite{Dawson94,SGY97}.

Dorfman's two outcomes are therefore not mutually exclusive alternatives. At any fixed noise level, random perturbations make a non-hyperbolic system behave as if it were hyperbolic with respect to its bulk statistics, yet they seriously complicate the tails and predictability, where they single out the very structures the bulk obscures. Table~\ref{tab:faces} summarizes what is currently established across the four attractor types. The empirical and numerical evidence remains uneven: it rests on a handful of model systems, stems largely from discrete-time maps, and is entirely absent for wild attractors.

\section{Consequences}
\label{sec:consequences}

The first consequence is practical: three recommendations follow for relating chaotic models to real physical systems. \emph{Declare the noise:} any chaotic model should incorporate an explicit stochastic term whose intensity is stated and exceeds the numerical noise of the integration scheme. \emph{Report ensembles:} a single trajectory should be accompanied by the spread of its ensemble, and any prediction horizon should be reported as a distribution. \emph{Judge features at the scale of the noise:} a periodic window, stable orbit, or bifurcation in a model that fails to survive noise at the level present in the physical system should not be attributed to that system.

The second consequence concerns structural error in the model, as distinct from uncertainty in the state. A numerical integration scheme introduces both, the former through truncation and the latter through round-off (Sec.~\ref{sec:nonoisefree}), and in this respect it mirrors the general physical situation, since the governing equations of a real system are never known exactly. Sensitivity to error in the initial state is the familiar butterfly effect; sensitivity to error in the model structure has been termed the hawkmoth effect~\cite{Frigg14}. I conjecture the following relation between the two. A trajectory of a slightly misspecified model constitutes a pseudo-orbit of the true system. For a hyperbolic attractor, shadowing therefore converts a small structural error in the model into an equivalent shift in the initial state, so that model error adds nothing to state error as far as statistics are concerned. Where hyperbolicity breaks down, however, this equivalence fails with it, and the two sources of error can no longer be separated. If this conjecture holds, the hawkmoth effect is localized at the non-hyperbolic structures of the attractor.

The third consequence applies to models learned directly from data. Any such model carries structural error by construction, and its training data are invariably noisy. A model trained on a single numerical trajectory learns the underlying system and the numerical integration scheme together~\cite{Khovanov21}. If the arguments developed in this paper hold, what a learned model can reproduce reliably is only what survives perturbation: the bulk statistics at the noise level of its training data and, I conjecture, the phase-space structures selected by large deviations, but never the deterministic fine structure of a quasi-attractor. The recent shift from training on individual trajectories to training on invariant statistics~\cite{Schiff24} aligns naturally with this conclusion. Two practical cautions follow. First, for a quasi-attractor, the stationary statistics depend on the noise intensity, so a model trained at a given noise level represents the physical system at that level alone. Second, both training and evaluation should rely on stochastic data and ensemble metrics, for the reasons set out in Sec.~\ref{sec:realisation}. In short, only the robust features of a chaotic system can be learned, and which features are robust depends on the noise.

\section{Open Problems}
\label{sec:open}

The view advanced here rests on a small number of model systems. Below, I outline the open problems whose resolution would confirm or refute it.

\begin{enumerate}
\item \emph{The four attractor types.} Does the threefold distinction between bulk statistics, large deviations, and predictability hold across all four types of attractor? For hyperbolic flows and for wild attractors, almost no results on the effects of noise exist (Table~\ref{tab:faces}).
\item \emph{The origin of the optimal path.} Does a large deviation from a non-hyperbolic attractor originate at a homoclinic tangency~\cite{KG04} or on a saddle cycle~\cite{Khovanov00,Khovanov08}? The two answers may pertain to distinct classes of systems, or one may prove to be a limiting case of the other.
\item \emph{Escape and predictability.} Do the phase-space regions where predictability is lost coincide with the structures that select the optimal escape path? In the Lorenz system, they appear to do so~\cite{Khovanov21,FNL01}. Whether this holds for a quasi-attractor remains unknown.
\item \emph{Robustness of the dynamical skeleton.} Do these selecting structures form a skeleton that persists, and varies continuously, when system parameters and noise intensity are varied within a Newhouse region, where the deterministic fine structure does not?
\item \emph{Equivalent noise.} Can the numerical noise of a given integration scheme and step size be represented faithfully by an explicit stochastic term, and for which statistics? An affirmative answer would endow every numerical simulation with a well-defined effective noise level. For the ensemble of numerical schemes in Fig.~\ref{fig:lorenz}, the prediction horizon matches that of a stochastic ensemble with $D$ of order $10^{-9}$, yet matching the horizon alone remains a weak test.
\item \emph{The resolution set by noise.} What scaling law relates the noise intensity to the resolution threshold, in both parameter space and phase space, below which the fine structure of a quasi-attractor is washed out? For the period-doubling cascade, this law is established~\cite{CNR81,CFH82}; for a Newhouse region it is not.
\end{enumerate}

These are problems for experiment as much as for theory. For example, an electronic circuit is a physical perturbation of its own mathematical model. Circuit noise levels can be controlled and measured, and all four types of chaotic attractors can now be realized in hardware or formulated explicitly as continuous flows~\cite{Kuznetsov05,GKT21}. The Andronov school originally posed robustness against noise and robustness against structural model error (roughness) as two distinct demands. For chaotic systems, the evidence assembled here suggests that they are ultimately one and the same, and that non-hyperbolicity is precisely where this unified demand must be met.

\section*{Acknowledgements}

The author wishes to express his sincere gratitude to Professor Laszlo Kish, former Editor-in-Chief, and Professor Peter McClintock, current Editor-in-Chief of \emph{Fluctuation and Noise Letters}, whose enduring support, encouragement, and trust have been invaluable throughout his career.

\end{document}